\documentclass[12pt]{article}
\usepackage{graphicx}
\usepackage{latexsym}
\usepackage{amscd}
\usepackage[cp1251]{inputenc}
\usepackage[english,russian]{babel}
\usepackage{amsmath,  eucal}
\usepackage{amssymb}
\usepackage{indentfirst}
\usepackage[small,centerlast]{caption2}
\usepackage{longtable}
\usepackage[dvipsnames]{color}
\usepackage{cite}
\graphicspath{{figure/}}


 {\renewcommand{\baselinestretch}{1.1}

\makeatletter

\renewcommand{\section}{\@startsection {section}{1}{\z@}%
                                   {-3.5ex \@plus -1ex \@minus -.2ex}%
                                   {2.3ex \@plus.2ex}%
                                   {\normalfont\Large\uppercase}}
\renewcommand{\subsection}{\@startsection{subsection}{2}{\z@}%
                                     {-3.25ex\@plus -1ex \@minus -.2ex}%
                                     {1.5ex \@plus .2ex}%
                                     {\normalfont\large\itshape}}
\renewcommand{\subsubsection}{\@startsection{subsubsection}{3}{1em}%
                                     {-3.25ex\@plus -1ex \@minus -.2ex}%
                                     {-1.5em \@plus .2em}%
                                     {\normalfont\normalsize\bfseries}}

\makeatother

\begin{document}

\newcommand{\mc}[1]{\mathcal{#1}}
\newcommand{\E}{\mc{E}}
\thispagestyle{empty} \large
\renewcommand{\abstractname}{}

 \begin{center}
{\Large \textbf{Nonlinear longitudinal current, generated by
two transversal electromagnetic waves in collisionless plasma}}
\medskip

{\large \textbf{A. V. Latyshev$^*$, A. A. Yushkanov$^*$,  O. D.
Algazin$^{**}$, A. V. Kopaev$^{**}$,  V. S. Popov$^{**}$ }}
\medskip

\textit{$^*$ Moscow State Regional University
\\$^{**}$ Bauman Moscow State Technical University}

\end{center}

\begin{abstract}
\noindent\small
Classical plasma with any degree degeneration of electronic gas
is considered. In plasma two external electromagnetic field are
propagation.
It is required to find the plasma response on
these fields.
From kinetic Vlasov equation for collisionless plasmas
distribution function  in square-law approximation on sizes
of intensivities of two electric fields is received.

The formula for calculation electric
current at any temperature (any degree of degeneration
electronic gas) is deduced.
This formula contains an one-dimensional quadrature.

It is shown, that the nonlinearity account leads to occurrence
the longitudinal electric current directed along a wave vector.
This longitudinal current is perpendicular to  known tranversal
classical current, received at the linear analysis.

The case of small values of wave number is considered.
Graphic comparison of dimensionless size of the current
depending on wave number and frequency of oscillation
of the electromagnetic fields is carry out.
\end{abstract}

\section*{ВВЕДЕНИЕ}

В настоящей работе выводятся формулы для вычисления электрического
тока в бесстолкновительной плазме Ферми---Дирака.

При решении кинетического уравнения Власова, описывающего
поведение плазмы, мы учитываем как в разложении
функции распределения, так и в разложении  величины самосогласованного
электромагнитного поля величины, пропорциональные
квадратам напряженностей внешних электрических полей и их произведению.

При таком нелинейном подходе оказывается, что электрический ток
имеет две ненулевые компоненты. Одна компонента электрического
тока направлена вдоль напряженности электрического поля.
Эта компонента электрического поля
точно такая же, как и в линейном анализе. Это "поперечный"\,
ток.
Таким образом, в линейном приближении мы получаем известное
выражение поперечного электрического тока.

Вторая ненулевая компонента электрического тока имеет второй
порядок малости относительно величин напряженностей
электрических полей.
Вторая компонента электрического тока направлена вдоль волнового
вектора. Этот ток
ортогонален первой компоненте. Это "продольный"\, ток.

Появление продольного тока выявляется проведенным
нелинейным анализом взаимодействия электромагнитных полей
с плазмой.

Нелинейные эффекты в  плазме изучаются уже длительное время
\cite{Gins}--\cite{Lat1}.

В работах \cite{Gins} и \cite{Zyt} изучаются нелинейные эффекты в
плазме. В работе \cite{Zyt} нелинейный ток использовался, в
частности, в вопросах вероятности распадных процессов. Отметим,
что в работе \cite{Zyt2} указывается на существование
нелинейного тока вдоль волнового вектора (см. формулу (2.9) из \cite{Zyt2}).

Квантовая плазма изучалась в работах \cite{Lat1}--\cite{Lat8}.
Столкновительная квантовая плазма начала изучаться в работе
Мермина \cite{Mermin}. Затем квантовая столкновительная плазма
изучалась в наших работах \cite{Lat2}--\cite{Lat5}. Нами
изучалась квантовая столкновительная плазма с переменной частотой
столкновений. В работах    \cite{Lat7} -- \cite{Lat9} было
исследовано генерирование продольного тока поперечным
электромагнитным полем в классической и квантовой плазме
Ферми---Дирака \cite{Lat7}, в максвелловской плазме
\cite{Lat8} и в вырожденной плазме \cite{Lat9}.

Укажем еще на ряд работ по  плазме, в том числе и  квантовой.
Это работы \cite{Andres}--\cite{Manf}.

В настоящей работе выводятся формулы для вычисления электрического
тока в бесстолкновительной плазме
при произвольной температуре, то-есть при произвольной
степени вырождения электронного газа.

\section{Уравнение Власова}

Покажем, что в случае классической плазмы, описываемой уравнением
Власова, генерируется продольный ток и вычислим его плотность. На
существование этого тока указывалось более полувека тому назад
\cite{Zyt2}.

Возьмем уравнение Власова, описывающее поведение
бесстолкновительной плазмы
$$
\dfrac{\partial f}{\partial t}+\mathbf{v}\dfrac{\partial f}{\partial
\mathbf{r}}+
e\bigg(\mathbf{E}_1+{\mathbf E}_2+
\dfrac{1}{c}[\mathbf{v},\mathbf{H}_1+{\mathbf H}_2]\bigg)
\dfrac{\partial f}{\partial\mathbf{p}}=0.
\eqno{(1.1)}
$$

В уравнении (1.1) $f$ -- функция распределения электронов
плазмы, ${\bf E}_j, {\bf H}_j\quad (j=1,2)$ -- компоненты
электромагнитного поля,
$c$ -- скорость света, ${\bf p}=m{\bf v}$ -- импульс электронов,
${\bf v}$ -- скорость электронов,
функция распределения электронов Ферми---Дирака,
$f^{(0)}=f_{eq}({\bf r},v)$ (eq$\equiv$ equi\-lib\-rium) --
-- локально равновесное распределение Ферми---Дирака,
$$
f_{eq}({\bf r},v)=\Big[1+\exp\dfrac{\E-\mu({\bf r})}{k_BT}\Big]^{-1}=
\big[1+\exp(P^2-\alpha({\bf r}))\big]^{-1}=f_{eq}({\bf r},P)
$$
$\E={mv^2}/{2}$ -- энергия электронов, $\mu$ -- химический
потенциал электронного газа, $k_B$ -- постоянная Больцмана, $T$
-- температура плазмы, ${\bf P}={{\bf P}}/{p_T}$ --
безразмерный импульс электронов,  $p_T=mv_T$,
$v_T$ -- тепловая скорость электронов, $v_T=\sqrt{2k_BT/m}$,
$\alpha=\mu/(k_BT)$ -- безразмерный химический потенциал,
$k_BT=\E_T=mv_T^2/2$ -- тепловая кинетическая энергия электронов.

Ниже нам понадобится абсолютное распределение Ферми---Дирака
$f_0(v)$,
$$
f_0(v)=\Big[1+\exp\dfrac{\E-\mu}{k_BT}\Big]^{-1}=
\big[1+\exp(P^2-\alpha)\big]^{-1}=f_0(P),
$$

Будем считать, что в плазме имеется электромагнитное поле,
представляющее собой бегущую гармоническую волну:
$$
{\bf E}_j={\bf E}_{0j}e^{i({\bf k_jr}-\omega_j t)}, \qquad
{\bf H}_j={\bf H}_{0j}e^{i({\bf k_jr}-\omega_j t)}, \quad j=1,2.
$$

Электрическое и магнитное поля связаны между собой через
векторный потенциал поля:
$$
\mathbf{E}_j=-\dfrac{1}{c}\dfrac{\partial \mathbf{A}_{j}}{\partial
t}=\dfrac{i\omega_j}{c}\mathbf{A}_j,\;\qquad
\mathbf{H}_j={\rm rot} \mathbf{A}_j,\quad j=1,2.
$$

Будем считать, что векторный потенциал электромагнитного поля
${\bf A}_j({\bf r},t)$ ортогонален волновому вектору ${\bf
k}_j$, т.е.
$$
{\bf k}_j\cdot {\bf A}_j({\bf r},t)=0,\qquad j=1,2.
$$
Это значит, что волновой вектор ${\bf k}_j$ ортогонален
электрическому и магнитному полям:
$$
{\bf k}_j\cdot {\bf E}_j({\bf r},t)=
{\bf k}_j\cdot {\bf H}_j({\bf r},t)=0,\qquad j=1,2.
$$

Для определенности будем считать, что волновые векторы обеих
полей направлены вдоль оси $x$, а электромагнитные поля
направлены вдоль оси $y$,
т.е.
$$
{\bf k}_j=k_j(1,0,0), \qquad {\bf E}_j=E_{j}(x,t)(0,1,0).
$$

Следовательно,
$$
{\bf E}_j=-\dfrac{1}{c}\dfrac{\partial {\bf  A}_j}{\partial t}=
\dfrac{i\omega}{c}{\bf A}_j,
$$
$$
{\bf H}_j=\dfrac{ck_j}{\omega}E_{j}\cdot(0,0,1),\qquad
{\bf [v,H}_j]=\dfrac{ck_j}{\omega_j}E_{j}\cdot (v_y,-v_x,0),
$$\medskip
$$
E_j=E_j(x,t)=E_{0j}e^{i(k_jx-\omega_j t)}\quad (j=1,2),
$$ \medskip
$$
e\bigg(\mathbf{E}_j+\dfrac{1}{c}[\mathbf{v},\mathbf{H}_j]\bigg)
\dfrac{\partial f}{\partial\mathbf{p}}=
\dfrac{e}{\omega_j}E_{j}\Big[k_jv_y\dfrac{\partial f}{\partial p_x}+
(\omega_j-k_jv_x)\dfrac{\partial f}{\partial p_y}\Big],
$$ \medskip
а также
$$
[{\bf v, H}_j]\dfrac{\partial f_0}{\partial {\bf p}}=0, \quad
\text{так как}\quad \dfrac{\partial f_0}{\partial {\bf p}}\sim
{\bf v}.
$$

Уравнение (1.1) может быть переписано в виде
$$
\dfrac{\partial f}{\partial t}+v_x\dfrac{\partial f}{\partial x}+
e\sum\limits_{j=1}^{2}\dfrac{E_j}{\omega_j}
\Big[k_jv_y\dfrac{\partial f}{\partial p_x}+
(\omega_j-k_jv_x)\dfrac{\partial f}{\partial p_y}\Big]=0.
\eqno{(1.2)}
$$

Будем искать решение уравнения (1.2) в виде
$$
f=f_0(P)+f_1+f_2,
\eqno{(1.3)}
$$
где
$$
f_1=E_1\varphi_1+E_2\varphi_2,
\eqno{(1.4)}
$$
$$
f_2=E_1^2\psi_1+E_2^2\psi_2+E_1E_2\psi_{0}.
\eqno{(1.5)}
$$

\section{Первое приближение}

Будем действовать методом последовательных приближений,
считая малым параметром величину напряженности электрического поля.
Тогда уравнение (1.2) с помощью (1.3) эквивалентно следующим уравнениям
$$
\dfrac{\partial f_1}{\partial t}+
v_x\dfrac{\partial f_1}{\partial x}=
-e\sum\limits_{j=1}^{2}\dfrac{E_j}{\omega_j}
\Big[k_jv_y\dfrac{\partial f_0}{\partial p_x}+
(\omega_j-k_jv_x)\dfrac{\partial f_0}{\partial p_y}\Big]
\eqno{(2.1)}
$$ \bigskip
и
$$
\dfrac{\partial f_2}{\partial t}+
v_x\dfrac{\partial f_2}{\partial x}=
-e\sum\limits_{j=1}^{2}\dfrac{E_j}{\omega_j}
\Big[k_jv_y\dfrac{\partial f_1}{\partial p_x}+
(\omega_j-k_jv_x)\dfrac{\partial f_1}{\partial p_y}\Big].
\eqno{(2.2)}
$$ \bigskip

Из уравнения (2.1) получаем
$$
[E_1(-i\omega_1+ik_1v_x)\varphi_1+E_2(-i\omega_2+ik_2v_x)\varphi_2]=
$$
$$
=-e\sum\limits_{j=1}^{2}\dfrac{E_j}{\omega_j}
\Big[k_jv_y\dfrac{\partial f_0}{\partial p_x}+
(\omega_j-k_jv_x)\dfrac{\partial f_0}{\partial p_y}\Big].
\eqno{(2.3)}
$$

Введем безразмерные параметры
$$
\Omega_j=\dfrac{\omega_j}{k_Tv_T},\qquad \qquad q_j=\dfrac{k_j}{k_T}.
$$

Здесь $q_j$ -- безразмерное волновое число,
$k_T =\dfrac {mv_T} {\hbar} $ -- тепловое волновое число, $ \Omega_j$
-- безразмерная частота колебаний электромагнитного поля ${\bf E}_j$.

В уравнении (2.3) перейдем к безразмерным параметрам. В
результате получаем уравнение
$$
i[E_1(q_1P_x-\Omega_1)\varphi_1+E_2(q_2P_x-\Omega_2)\varphi_2]=
$$
$$
=-\dfrac{e}{k_Tp_Tv_T}\sum\limits_{j=1}^{2}\dfrac{E_j}{\Omega_j}
\Big[q_jP_y\dfrac{\partial f_0}{\partial P_x}+
(\Omega_j-q_jP_x)\dfrac{\partial f_0}{\partial P_y}\Big].
\eqno{(2.4)}
$$

Заметим, что
$$
\dfrac{\partial f_0}{\partial P_x}\sim P_x,\qquad
\dfrac{\partial f_0}{\partial P_y}\sim P_y.
$$

Следовательно
$$
\Bigg[q_jP_y\dfrac{\partial f_0}{\partial P_x}+
(\Omega_j-q_jP_x)\dfrac{\partial f_0}{\partial P_y}\Bigg]=
\Omega_j\dfrac{\partial f_0}{\partial P_y}
$$
и уравнение (2.4) упрощается:
$$
i[E_1(q_1P_x-\Omega_1)\varphi_1+E_2(q_2P_x-\Omega_2)\varphi_2]=
$$
$$
=-\dfrac{e}{k_Tp_Tv_T}(E_1+E_2)\dfrac{\partial f_0}{\partial P_y}.
\eqno{(2.5)}
$$
Из уравнения (2.5) находим, что
$$
\varphi_j=\dfrac{ie}{k_Tp_Tv_T}\cdot
\dfrac{{\partial f_0}{\partial P_y}}{q_jP_x-\Omega_j},\qquad
j=1,2.
\eqno{(2.6)}
$$

Теперь из уравнения (2.6) находим, что
$$
f_1=\dfrac{ie}{k_Tp_Tv_T}\cdot\Bigg[\dfrac{E_1}{q_1P_x-\Omega_1}+
\dfrac{E_2}{q_2P_x-\Omega_2}\Bigg]\dfrac{\partial f_0}{\partial P_y}.
\eqno{(2.7)}
$$

\section{Второе приближение}

Перейдем ко второму приближению.
В левую часть уравнения (2.2) подставим (1.5).
$$
f_2=E_{1}^2\psi_1+E_{2}^2\psi_2+E_{1}E_{2}\psi_0.
$$

Получим уравнение
$$
\big\{E_1^2(-2i\omega_1+2ik_1v_x)\psi_1+E_2^2(-2i\omega_2+2ik_2v_x)\psi_2+
$$$$+E_1E_2[-i(\omega_1+\omega_2)+i(k_1+k_2)v_x]\psi_0\big\}=
$$
$$
=-\dfrac{eE_1}{\omega_1}
\Big[k_1v_y\dfrac{\partial f_1}{\partial p_x}+
(\omega_1-k_1v_x)\dfrac{\partial f_1}{\partial p_y}\Big]
-\dfrac{eE_2}{\omega_2}
\Big[k_2v_y\dfrac{\partial f_1}{\partial p_x}+
(\omega_2-k_2v_x)\dfrac{\partial f_1}{\partial p_y}\Big].
$$

Перейдем в этом уравнении к безразмерным параметрам и введем
обозначения
$$
q=\dfrac{q_1+q_2}{2}, \qquad
\Omega=\dfrac{\Omega_1+\Omega_2}{2}.
$$

Получим уравнение
$$
2i\Big\{E_1^2(q_1P_x-\Omega_1)\psi_0+E_2^2(q_2-\Omega_2)\psi_2+
E_1E_2(qP_x-\Omega)\psi_0\Big\}=
$$
$$
=-\dfrac{ie^2}{k_T^2p_T^2v_T^2}\cdot\dfrac{E_1}{\Omega_1}\cdot
\Bigg\{q_1P_y\dfrac{\partial}{\partial P_x}
\Bigg[\Big(\dfrac{E_1}{q_1P_x-\Omega_1}+\dfrac{E_2}{q_2P_x-\Omega_2}\Big)
\cdot \dfrac{\partial f_0}{\partial P_y}\Bigg]+
$$$$+
(\Omega_1-q_1P_x)\dfrac{\partial }
{\partial P_y}\Bigg[\Big(\dfrac{E_1}{q_1P_x-\Omega_1}+
\dfrac{E_2}{q_2P_x-\Omega_2}\Big)\cdot
\dfrac{\partial f_0}{\partial P_y}\Bigg]\Bigg\}-
$$
$$
-\dfrac{ie^2}{k_T^2p_T^2v_T^2}\cdot\dfrac{E_2}{\Omega_2}\cdot
\Bigg\{q_2P_y\dfrac{\partial}{\partial P_x}
\Bigg[\Big(\dfrac{E_1}{q_1P_x-\Omega_1}+\dfrac{E_2}{q_2P_x-\Omega_2}\Big)
\cdot \dfrac{\partial f_0}{\partial P_y}\Bigg]+
$$
$$+
(\Omega_2-q_2P_x)\dfrac{\partial }
{\partial P_y}\Bigg[\Big(\dfrac{E_1}{q_1P_x-\Omega_1}+
\dfrac{E_2}{q_2P_x-\Omega_2}\Big)
\cdot \dfrac{\partial f_0}{\partial P_y}\Bigg]\Bigg\}.
$$

Преобразуем предыдущее уравнение
$$
E_1^2(q_1P_x-\Omega_1)\psi_0+E_2^2(q_2-\Omega_2)\psi_2+
E_1E_2(qP_x-\Omega)\psi_0=
$$
$$
=-\dfrac{e^2}{2k_T^2p_T^2v_T^2}\Bigg\{\dfrac{E_1^2}{\Omega_1}
\Bigg[q_1P_y\dfrac{\partial}{\partial P_x}\Big(\dfrac{\partial f_0
/\partial P_y}{q_1P_x-\Omega_1}\Big)-\dfrac{\partial^2 f_0}
{\partial P_y^2}\Bigg]+
$$
$$
+\dfrac{E_2^2}{\Omega_2}
\Bigg[q_2P_y\dfrac{\partial}{\partial P_x}\Big(\dfrac{\partial f_0
/\partial P_y}{q_2P_x-\Omega_2}\Big)-\dfrac{\partial^2 f_0}
{\partial P_y^2}\Bigg]+
$$
$$
+\dfrac{E_1E_2}{\Omega_1}
\Bigg[q_1P_y\dfrac{\partial}{\partial P_x}\Big(\dfrac{\partial f_0
/\partial P_y}{q_2P_x-\Omega_2}\Big)+\dfrac{\Omega_1-q_1P_x}
{q_2P_x-\Omega_2}\cdot\dfrac{\partial^2 f_0}{\partial P_y^2}\Bigg]+
$$
$$
+\dfrac{E_1E_2}{\Omega_2}
\Bigg[q_2P_y\dfrac{\partial}{\partial P_x}\Big(\dfrac{\partial f_0
/\partial P_y}{q_1P_x-\Omega_1}\Big)+\dfrac{\Omega_2-q_2P_x}
{q_1P_x-\Omega_1}\cdot\dfrac{\partial^2 f_0}{\partial
P_y^2}\Bigg]\Bigg\}.
$$

Из этого уравнения находим
$$
\psi_1=-\dfrac{e^2}{2k_T^2p_T^2v_T^2\Omega_1}\cdot
\dfrac{\Xi_1({\bf P})}{q_1P_x-\Omega_1},
\eqno{(3.1)}
$$
$$
\psi_2=-\dfrac{e^2}{2k_T^2p_T^2v_T^2\Omega_2}\cdot
\dfrac{\Xi_2({\bf P})}{q_2P_x-\Omega_2},
\eqno{(3.2)}
$$
$$
\psi_0=-\dfrac{e^2}{2k_T^2p_T^2v_T^2}\cdot
\Bigg[
\dfrac{1}{\Omega_1}\cdot\dfrac{\Xi_{12}({\bf P})}{qP_x-\Omega}+
\dfrac{1}{\Omega_2}\cdot\dfrac{\Xi_{21}({\bf P})}{qP_x-\Omega}\Bigg],
\eqno{(3.3)}
$$

где
$$
\Xi_j({\bf P})=q_jP_y\dfrac{\partial}{\partial P_x}
\Big(\dfrac{\partial f_0/\partial P_y}{q_jP_x-\Omega_1}\Big)-
\dfrac{\partial^2 f_0}{\partial P_y^2},
$$

$$
\Xi_{12}({\bf P})=q_1P_y\dfrac{\partial}{\partial P_x}
\Big(\dfrac{\partial f_0/\partial P_y}{q_2P_x-\Omega_2}\Big)+
\dfrac{\Omega_1-q_1P_x}{q_2P_x-\Omega_2}\cdot
\dfrac{\partial^2 f_0}{\partial P_y^2},
$$

$$
\Xi_{21}({\bf P})=q_2P_y\dfrac{\partial}{\partial P_x}
\Big(\dfrac{\partial f_0/\partial P_y}{q_1P_x-\Omega_1}\Big)+
\dfrac{\Omega_2-q_2P_x}{q_1P_x-\Omega_1}\cdot
\dfrac{\partial^2 f_0}{\partial P_y^2}.
$$

Функция распределения во втором приближении по полю построена и
определяется равенством (1.3), в котором функция $f_1$
определяется равенством (1.4), а функция $f_2$ определяется
равенствами (1.5) и (3.1)--(3.3).

\section{Плотность поперечного электрического\\ тока}

Найдем плотность электрического тока
$$
\mathbf{j}=e\int \mathbf{v}f \dfrac{2d^3p}{(2\pi\hbar)^3}=
\dfrac{2p_T^3v_T}{(2\pi\hbar)^3}\int f{\bf P}d^3P.
\eqno{(4.1)}
$$
Из равенств (1.3)--(1.5) и (4.1) видно, что вектор плотности тока имеет
две ненулевые компоненты
$$
\mathbf{j}=(j_x,j_y,0).
$$

Здесь $j_y$ -- плотность поперечного тока, $j_x$ -- плотность
продольного тока,
$$
j_y=e\int v_yf \dfrac{2d^3p}{(2\pi\hbar)^3}=
e\int v_yf_1 \dfrac{2d^3p}{(2\pi\hbar)^3}=
\dfrac{e2p_T^3v_T}{(2\pi\hbar)^3}\int f_1 P_yd^3P.
\eqno{(4.2)}
$$
$$
j_x=e\int v_yf \dfrac{2d^3p}{(2\pi\hbar)^3}=
e\int v_yf_2 \dfrac{2d^3p}{(2\pi\hbar)^3}=
\dfrac{e2p_T^3v_T}{(2\pi\hbar)^3}\int f_2 P_yd^3P.
\eqno{(4.3)}
$$

Согласно (4.2) в явном виде поперечный ток равен:
$$
j_y=\dfrac{2ie^2p_T^3}{(2\pi\hbar)^2k_Tp_T}\int \Big(\dfrac{E_1}
{q_1P_x-\Omega_1}+\dfrac{E_2}{q_2P_x-\Omega_2}\Big)
\dfrac{\partial f_0}{\partial P_y}P_y d^3P.
\eqno{(4.4)}
$$

Этот ток направлен вдоль электромагнитного поля, его плотность
определяется только первым приближением функции распределения.

Второе приближение функции распределения вклад в плотность
поперечного тока не вносит.

Упростим формулу (4.4). Заметим, что внутренний интеграл по
переменной $P_y$ равен:
$$
\int\limits_{-\infty}^{\infty}
P_y\dfrac{\partial f_0(P)}{\partial P_y}dP_y=$$$$=
P_yf_0(P)\Bigg|_{P_y=-\infty}^{P_y=+\infty}-\int\limits_{-\infty}^{\infty}
f_0(P)dP_y=-\int\limits_{-\infty}^{\infty}
f_0(P)dP_y.
\eqno{(4.5)}
$$

Следовательно, поперечный ток равен:
$$
j_y=-\dfrac{2ie^2p_T^2}{(2\pi\hbar)^3k_T}
\int \Bigg(\dfrac{E_1}
{q_1P_x-\Omega_1}+\dfrac{E_2}{q_2P_x-\Omega_2}\Bigg)
f_0(P)d^3P.
$$

Внутренний интеграл в плоскости $(P_y,P_z)$ вычислим в полярных
координатах:
$$
\int\limits_{-\infty}^{\infty}\int\limits_{-\infty}^{\infty}
f_0(P)dP_ydP_z=\pi\ln(1+e^{\alpha-P_x^2}).
\eqno{(4.6)}
$$

Тогда получаем, что
$$
j_y=-\dfrac{2\pi ie^2p_T^2}{(2\pi\hbar)^3k_T}
\int\limits_{-\infty}^{\infty}\Bigg(\dfrac{E_1}
{q_1P_x-\Omega_1}+\dfrac{E_2}{q_2P_x-\Omega_2}\Bigg)
\ln(1+e^{\alpha-P_x^2})dP_x.
\eqno{(4.7)}
$$

\section{Плотность продольного электрического\\ тока}

Перейдем к исследованию продольного тока (4.3).

С помощью разложения (1.5) представим продольный ток в виде трех
слагаемых:
$$
j_x=j_1+j_2+j_0.
\eqno{(5.1)}
$$

Здесь
$$
j_1=E_1^2\dfrac{2ep_Tv_T}{(2\pi \hbar)^3}\int P_x\psi_1d^3P,
$$
$$
j_2=E_2^2\dfrac{2ep_Tv_T}{(2\pi \hbar)^3}\int P_x\psi_2d^3P,
$$
$$
j_0=E_1E_2\dfrac{2ep_Tv_T}{(2\pi \hbar)^3}\int P_x\psi_0d^3P.
$$

Представим составляющие продольного тока равенства в явном виде:
$$
j_1=-\dfrac{e^3p_TE_1^2}{(2\pi\hbar)^3k_T^2v_T\Omega_1}\int
\dfrac{\Xi_1({\bf P})P_xd^3P}{q_1P_x-\Omega_1},
$$

$$
j_1=-\dfrac{e^3p_TE_2^2}{(2\pi\hbar)^3k_T^2v_T\Omega_2}\int
\dfrac{\Xi_2({\bf P})P_xd^3P}{q_2P_x-\Omega_2},
$$

$$
j_0=-\dfrac{e^3p_TE_1E_2}{(2\pi\hbar)^3k_T^2v_T}\int
\Bigg[\dfrac{1}{\Omega_1}\dfrac{\Xi_{12}({\bf
P})}{qP_x-\Omega}+\dfrac{1}{\Omega_2}\dfrac{\Xi_{21}({\bf
P})}{qP_x-\Omega}\Bigg]P_xd^3P.
\eqno{(5.4)}
$$

Заметим, что в этих выражениях одномерный внутренний интеграл по $P_y$
равен нулю:
$$
\int\limits_{-\infty}^{\infty}\dfrac{\partial^2f_0}{\partial P_y^2}dP_y
=\dfrac{\partial f_0}{\partial P_y}\Bigg|_{P_y=-\infty}^{P_y=+\infty}=0,
$$
а внутренние интегралы по $P_x$ вычисляются по частям:
$$
\int\limits_{-\infty}^{\infty}\dfrac{\partial}{\partial P_x}
\Big(\dfrac{\partial f_0/\partial P_y}{q_jP_x-\Omega_j}\Big)
\dfrac{P_xdP_x}{q_jP_x-\Omega_j}=\Omega_j\int\limits_{-\infty}^{\infty}
\dfrac{[\partial f_0/\partial P_y]dP_x}{(q_jP_x-\Omega_j)^3},
$$
где $j=1,2$, и
$$
\int\limits_{-\infty}^{\infty}\dfrac{\partial}{\partial P_x}
\Big(\dfrac{\partial f_0/\partial P_y}{q_jP_x-\Omega_j}\Big)
\dfrac{P_xdP_x}{qP_x-\Omega}=\Omega\int\limits_{-\infty}^{\infty}
\dfrac{[\partial f_0/\partial P_y]dP_x}{(q_jP_x-\Omega_j)(qP_x-\Omega)^2}.
$$

Следовательно, предыдущие равенства для составляющих продольного
тока упрощаются:
$$
j_1=-\dfrac{e^3E_1^2p_Tq_1}{(2\pi\hbar)^3k_T^2v_T}\int
\dfrac{P_y(\partial f_0/\partial P_y)d^3P}{(q_1P_x-\Omega_1)^3},
$$
$$
j_2=-\dfrac{e^3E_1^2p_Tq_2}{(2\pi\hbar)^3k_T^2v_T}\int
\dfrac{P_y(\partial f_0/\partial P_y)d^3P}{(q_2P_x-\Omega_2)^3},
$$
$$
j_0=j_{12}+j_{21},
$$
$$
j_{12}=-\dfrac{e^3E_1E_2p_Tq_1\Omega}{(2\pi\hbar)^3k_T^2v_T\Omega_1}\int
\dfrac{P_y(\partial f_0/\partial P_y)d^3P}
{(q_2P_x-\Omega_2)(qP_x-\Omega)^2},
$$
$$
j_{21}=-\dfrac{e^3E_1E_2p_Tq_2\Omega}{(2\pi\hbar)^3k_T^2v_T\Omega_2}\int
\dfrac{P_y(\partial f_0/\partial P_y)d^3P}
{(q_1P_x-\Omega_1)(qP_x-\Omega)^2}.
$$

Внутренний интеграл по переменной $P_y$  проинтегрируем по
частям и воспользуемся равенством (4.5). В результате получаем
следующие выражения для составляющих слагаемых
продольного тока:
$$
j_1=\dfrac{e^3E_1^2p_Tq_1}{(2\pi\hbar)^3k_T^2v_T}\int
\dfrac{f_0(P)d^3P}{(q_1P_x-\Omega_1)^3},
$$
$$
j_2=\dfrac{e^3E_1^2p_Tq_2}{(2\pi\hbar)^3k_T^2v_T}\int
\dfrac{f_0(P)d^3P}{(q_2P_x-\Omega_2)^3},
$$
$$
j_0=j_{12}+j_{21},
$$
$$
j_{12}=\dfrac{e^3E_1E_2p_Tq_1\Omega}{(2\pi\hbar)^3k_T^2v_T\Omega_1}\int
\dfrac{f_0(P)d^3P}
{(q_2P_x-\Omega_2)(qP_x-\Omega)^2},
$$
$$
j_{21}=\dfrac{e^3E_1E_2p_Tq_2\Omega}{(2\pi\hbar)^3k_T^2v_T\Omega_2}\int
\dfrac{f_0(P)d^3P}
{(q_1P_x-\Omega_1)(qP_x-\Omega)^2}.
$$

Внутренние интегралы в плоскости $(P_y,P_z)$ вычислим в полярных
координатах и воспользуемся формулой (4.6).
В результате предыдущие равенства сводятся к одномерному интегралу:
$$
j_1=\dfrac{\pi e^3E_1^2p_Tq_1}{(2\pi\hbar)^3k_T^2v_T}
\int\limits_{-\infty}^{\infty}
\dfrac{\ln(1+e^{\alpha-P_x^2})dP_x}
{(q_1P_x-\Omega_1)^3},
$$
$$
j_2=\dfrac{\pi e^3E_1^2p_Tq_2}{(2\pi\hbar)^3k_T^2v_T}
\int\limits_{-\infty}^{\infty}
\dfrac{\ln(1+e^{\alpha-P_x^2})dP_x}
{(q_2P_x-\Omega_2)^3},
$$
$$
j_0=j_{12}+j_{21},
$$
$$
j_{12}=\dfrac{\pi e^3E_1E_2p_Tq_1\Omega}{(2\pi\hbar)^3k_T^2v_T\Omega_1}
\int\limits_{-\infty}^{\infty}
\dfrac{\ln(1+e^{\alpha-P_x^2})dP_x}
{(q_2P_x-\Omega_2)(qP_x-\Omega)^2},
$$
$$
j_{21}=\dfrac{\pi e^3E_1E_2p_Tq_2\Omega}{(2\pi\hbar)^3k_T^2v_T\Omega_2}
\int\limits_{-\infty}^{\infty}\dfrac{\ln(1+e^{\alpha-P_x^2})dP_x}
{(q_1P_x-\Omega_1)(qP_x-\Omega)^2}.
$$

Найдем числовую плотность концентрацию частиц плазмы, отвечающую
распределению Ферми---Дирака
$$
N=\int f_0(P)\dfrac{2d^3p}{(2\pi\hbar)^3}=
\dfrac{8\pi p_T^3}{(2\pi\hbar)^3}\int\limits_{0}^{\infty}
\dfrac{e^{\alpha-P^2}P^2dP}{1+e^{\alpha-P^2}}=
\dfrac{k_T^3}{2\pi^2}l_0(\alpha),
$$
где
$$
l_0(\alpha)=\int\limits_{0}^{\infty}\ln(1+e^{\alpha-\tau^2})d\tau.
$$

В выражении перед интегралами выделим плазменную
(ленгмюровскую) частоту
$$
\omega_p=\sqrt{\dfrac{4\pi e^2N}{m}}
$$
и числовую плотность (концентрацию) $N$,
причем последнюю выразим через тепловое волновое число. Получим
$$
\dfrac{\pi p_T e^3q_j}{(2\pi\hbar)^3k_T^2v_T}=
\dfrac{e^3p_Tq_jN 2\pi^2}{8\pi^3k_T^2v_Tp_T^3l_0(\alpha)}=
$$
$$
=\dfrac{e\omega_p^2mq_jv_T}{16\pi k_T^2v_T^2p_T^2l_0(\alpha)}=
\dfrac{e\Omega_p^2}{p_Tk_T}\cdot\dfrac{k_j}{16\pi l_0(\alpha)}=
\sigma_{l,tr}\cdot\dfrac{k_j}{16\pi l_0(\alpha)}.
$$

Здесь
$$
\Omega_p=\dfrac{\omega_p}{k_Tv_T}=\dfrac{\hbar\omega_p}{mv_T^2}
$$
-- безразмерная плазменная частота,
$\sigma_{l,tr}$ -- продольно-поперечная проводимость,
$$
\sigma_{l,tr}=\dfrac{e\hbar}{p_T^2}
\Big(\dfrac{\hbar \omega_p}{mv_T^2}\Big)^2=
\dfrac{e}{k_Tp_T}\Big(\dfrac{\omega_p}{k_Tv_T}\Big)^2=
\dfrac{e\Omega_p^2}{p_Tk_T}.
$$

Теперь составляющие продольного тока записыватся в виде:
$$
j_1=E_1^2\sigma_{l,tr}k_1J_1,
\eqno{(5.2)}
$$
где
$$
J_1=\dfrac{1}{16\pi l_0(\alpha)}
\int\limits_{-\infty}^{\infty}
\dfrac{\ln(1+e^{\alpha-P_x^2})dP_x}
{(q_1P_x-\Omega_1)^3},
$$
$$
j_2=E_2^2\sigma_{l,tr}k_2J_2,
\eqno{(5.3)}
$$
где
$$
J_2=\dfrac{1}{16\pi l_0(\alpha)}
\int\limits_{-\infty}^{\infty}
\dfrac{\ln(1+e^{\alpha-P_x^2})dP_x}
{(q_2P_x-\Omega_2)^3},
$$
$$
j_0=E_1E_2 \sigma_{l,tr}(k_1J_{12}+k_2J_{21}),
\eqno{(5.4)}
$$
где
$$
J_{12}=\dfrac{\Omega}{16\pi l_0(\alpha)\Omega_1}
\int\limits_{-\infty}^{\infty}
\dfrac{\ln(1+e^{\alpha-P_x^2})dP_x}
{(q_2P_x-\Omega_2)(qP_x-\Omega)^2},
$$
$$
J_{21}=\dfrac{\Omega}{16\pi l_0(\alpha)\Omega_2}
\int\limits_{-\infty}^{\infty}\dfrac{\ln(1+e^{\alpha-P_x^2})dP_x}
{(q_1P_x-\Omega_1)(qP_x-\Omega)^2}.
$$

В равенствах (5.2)--(5.4) величины $J_1,J_2,J_{12},J_{21}$ --
безразмерные части плотности продольного тока.

Таким образом, продольная часть тока окончательно  равна:
$$
j_x=\sigma_{l,tr}[E_1^2k_1J_1+E_2^2k_2J_2+E_1E_2(k_1J_{12}+k_2J_{21})].
\eqno{(5.5)}
$$

Если ввести поперечные поля
$$
\mathbf{E}_j^{\bf \rm tr}=\mathbf{E}_j-
\dfrac{\mathbf{k}_j({\bf E}_j{\bf k}_j)}{k_j^2}=
\mathbf{E}_j-\dfrac{\mathbf{q}_j({\bf E}_j{\bf q}_j)}{q_j^2},
$$
то
равенство (5.5) можно записать в инвариантой форме
$$
{\bf j}_{long}=\sigma_{l,tr}[({\bf E}_1^{tr})^2{\bf k}_1J_1+
({\bf E}_2^{tr})^2{\bf k}_2J_2+{\bf E}_1^{tr}{\bf E}_2^{tr}
({\bf k}_1J_{12}+{\bf k}_2J_{21})].
$$

Перейдем к рассмотрению случая малых значений волнового числа. Из
выражениий (5.2)--(5.4) вытекает, что при малых значениях
волновых  чисел для плотности продольного тока получаем:
$$
j_x=-\dfrac{\sigma_{l,tr}}{8\pi}\Big[E_1^2\dfrac{k_1}{\Omega_1^3}+
E_2^2\dfrac{k_2}{\Omega_2^3}+2E_1E_2\dfrac{k_1+k_2}
{\Omega_1\Omega_2(\Omega_1+\Omega_2)}\Big].
$$

{\sc Замечание 1.} При вычислении интегралов, входящих в
безразмерные части плотности продольного тока, следует
воспользоваться известным правилом Ландау. Согласно правилу
Ландау для интегралов $J_1$ и $J_2$ имеем:
$$
J_j=\dfrac{1}{16\pi l_0(\alpha)}\int\limits_{-\infty}^{\infty}
\dfrac{\ln(1+e^{\alpha-\tau^2})d\tau}{(q_j\tau-\Omega_j)^3}=
$$
$$
=\dfrac{1}{16\pi l_0(\alpha)}\Bigg[
-i\dfrac{\pi}{2q_j^3}\Big[\ln(1+e^{\alpha-\tau^2})\Big]''
\Bigg|_{\tau=\Omega_j/q_j}+$$$$+{\rm V.p.}\int\limits_{-\infty}^{\infty}
\dfrac{\ln(1+e^{\alpha-\tau^2})d\tau}{(q_j\tau-\Omega_j)^3}\Bigg].
\eqno{(5.6)}
$$\medskip

Символ ${\rm V.p.}$ перед интегралом означает, что интеграл
понимается в смысле главного значения.

Чтобы воспользоваться правилом Ландау для интегралов $J_{12}$ и
$J_{21}$ следует разложить подынтегральные дроби на
элементарные. Нетрудно вычислить, что
$$
\dfrac{1}{(q_2\tau-\Omega_2)(q \tau-\Omega)^2}=\dfrac{a_1}{q_2-\tau-\Omega_2}+
\dfrac{b_1}{q\tau-\Omega}+\dfrac{c_1}{(q\tau-\Omega)^2},
\eqno{(5.7)}
$$
где
$$
a_1=\dfrac{4q_2^2}{(q_1\Omega_2-q_2\Omega_1)^2},
$$
$$
b_1=-\dfrac{4q_1q_2}{(q_1\Omega_2-q_2\Omega_1)^2},
$$
$$
c_1=-\dfrac{2q}{q_1\Omega_2-q_2\Omega_1}.
$$
Аналогично,
$$
\dfrac{1}{(q_1\tau-\Omega_1)(q \tau-\Omega)^2}=\dfrac{a_2}{q_2-\tau-\Omega_2}+
\dfrac{b_2}{q\tau-\Omega}+\dfrac{c_2}{(q\tau-\Omega)^2},
\eqno{(5.7)}
$$
где
$$
a_2=\dfrac{4q_1^2}{(q_1\Omega_2-q_2\Omega_1)^2},
$$
$$
b_2=-\dfrac{4q_1q_2}{(q_1\Omega_2-q_2\Omega_1)^2},
$$
$$
c_2=\dfrac{2q}{q_1\Omega_2-q_2\Omega_1}.
$$

На основании приведенных разложений интегралы $J_{12}$ и
$J_{21}$  согласно правилу Ландау вычисляются следующим образом:
$$
J_{12}=\dfrac{\Omega}{16\pi l_0(\alpha)\Omega_1}\Bigg[{\rm V.P.}
\int\limits_{-\infty}^{\infty}
\dfrac{\ln(1+e^{\alpha-\tau^2})d\tau}{(q_2\tau-\Omega_2)(q\tau-\Omega)^2}-
$$
$$
-i\pi a_1\ln(1+e^{\alpha-\Omega_2^2/q_2^2})-
i\pi b_1\ln(1+e^{\alpha-\Omega^2/q^2})+
2i\pi c_1\dfrac{\Omega}{q}
\dfrac{e^{\alpha-\Omega^2/q^2}}{1+e^{\alpha-\Omega^2/q^2}}\Bigg],
$$
и
$$
J_{21}=\dfrac{\Omega}{16\pi l_0(\alpha)\Omega_1}\Bigg[{\rm V.P.}
\int\limits_{-\infty}^{\infty}
\dfrac{\ln(1+e^{\alpha-\tau^2})d\tau}{(q_1\tau-\Omega_1)(q\tau-\Omega)^2}-
$$
$$
-i\pi a_2\ln(1+e^{\alpha-\Omega_1^2/q_1^2})-
i\pi b_2\ln(1+e^{\alpha-\Omega^2/q^2})+
2i\pi c_2\dfrac{\Omega}{q}
\dfrac{e^{\alpha-\Omega^2/q^2}}{1+e^{\alpha-\Omega^2/q^2}}\Bigg].
$$

\section{ЗАКЛЮЧЕНИЕ}

Проведем графическое исследование действительных и мнимых
частей  плотностей безразмерных "перекрестных"\, токов  $J_{12}$ и $J_{21}$.
В случае одного электромагнитного поля в работе \cite{Lat7} было
проведено графическое исследование безразмерной части
продольного тока. При этом исследовались составляющие
безразмерных плотностей тока вида $J_1$ (или $J_2$).
При этом использовались формулы (5.6). Поэтому в
настоящей работе будем исследовать "перекрестные"\,
величины $J_{12}$ и $J_{21}$.

На рис. 1, 2 и 5, 6 представлено поведение "перекрестных"\,токов
при значениях величины безразмерного химического потенциала
$\alpha=2, 0,-5$ в зависимости от безразмерного волнового числа $q_1$,
при этом $\Omega_1=1, \Omega_2=0.5, q_2=0.1$.

На рис. 1 и 2 представлено поведение
действительных частей плотностей
$J_{12}$ (рис. 1) и $J_{21}$ (рис. 2) , а на рис. 5 и 6 -- мнимых.

На рис. 3 и 4 представим поведение действительных частей
плотностей $J_{12}$ (рис. 3) и $J_{21}$ (рис. 4) при
$\Omega_1=1$, $q_2=0.1$ и при  $\Omega_2=0.1, 0.2, 0.3$
в зависимости от безразмерного волнового числа $q_1$. На рис. 7
и 8 представим поведение мнимых частей.

Из  рисунков 1--4 видно, что плотности "перекрестных"\, токов
имеют один минимум и один максимум, причем с ростом $q_1$ эти
величины стремятся к нулю.
С ростом величины химического потенциала (рис. 1 и 2) величина
максимума увеличивается, а минимума -- уменьшается.
С ростом частоты колебаний второго электромагнитного поля
величина максимума уменьшается, а величина минимума
увеличивается.

Из рис. 5 и 6 видно, что мнимая часть "перекрестных"\, токов
имеет резко выраженный максимум, минимум же наблюдается при
переходе к положительным значениям химического потенциала, т.е. с ростом
степени вырожденности электронного газа.
При этом независимо от величины химического потенциала и при
малых и при больших значениях волнового числа $q_1$ значения
мнимых частей плотностей "перекрестных"\, токов сближаются и в
пределе (при $q_1\to 0$ и $q_1\to \infty$) совпадают. С ростом
химического потенциала величина максимума увеличивается, а
минимума -- уменьшается.

Из рис. 7 и 8 видно, что с ростом волнового числа $q_1$ мнимые
части $J_{12}$ независимо от частоты колебаний второго
электромагнитного поля $\Omega_2$ сближаются и при $q_1\to \infty$
совпадают. Для плотности $J_{21}$ мнимые части всех кривых
совпадают и при малых и при больших значениях первого волнового
числа.

Отметим также, что мнимые части "перекрестных"\, токов имеют
максимум при всех значениях частоты колебаний второго
электромагнитного поля. При уменьшении частоты колебаний первого
электромагнитного поля у мнимых частей появляется минимум. При
уменьшении частоты колебаний $\Omega_1$ величина максимума
увеличивается, а минимума -- уменьшается.

В настоящей работе решена следующая задача: в плазме с произвольной
степенью вырождения электронного газа, распространяются две
электромагнитные волны с коллинеарными волновыми векторами.
Уравнение Власова решается методом последовательных приближений,
считая малыми параметрами одного порядка величины напряженностей
соответствующих электрических полей. Используется квадратичное разложение
функции распределения.

Оказалось, что учет нелинейности электромагнитных полей обнаруживает
генерирование электрического тока, ортогонального к направлению
электрического поля (т.е. направлению известного классического поперечного
электрического тока). Найдена величина поперечного и продольного
электрических токов.  Рассмотрен случай малых значений волновых чисел.
Проведено графическое исследование так называемых
"перекрестных"\, слагаемых $J_{12}$ и $J_{21}$, составляющих величину плотности
продольного электрического тока.

В дальнейшем авторы намерены рассмотреть задачи о колебаниях
плазмы и о скин-эффекте с использованием квадратичного по
потенциалу разложения функции распределения.

\clearpage

\begin{figure}[ht]\center
\includegraphics[width=16.0cm, height=10cm]{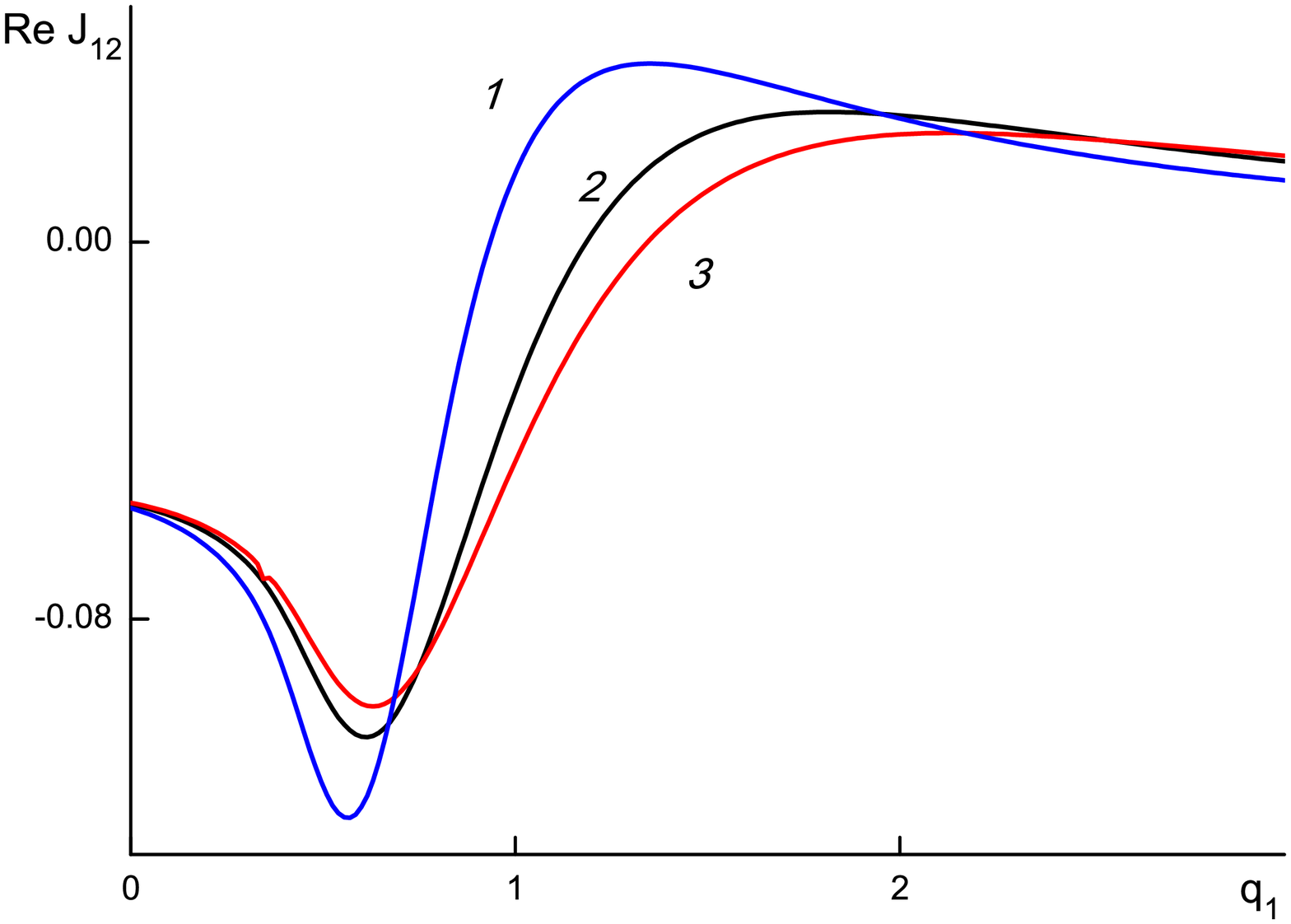}
\center{Fig. 1. Real part of dimensionless density of longitudinal
"crossed"\, current $J_{12}$, $\Omega_1=1, \Omega_2=0.5, q_2=0.1$.
Curves $1,2,3$ correspond to values of
dimensionless chemical potential $\alpha=2, 0, -5$.}
\includegraphics[width=17.0cm, height=10cm]{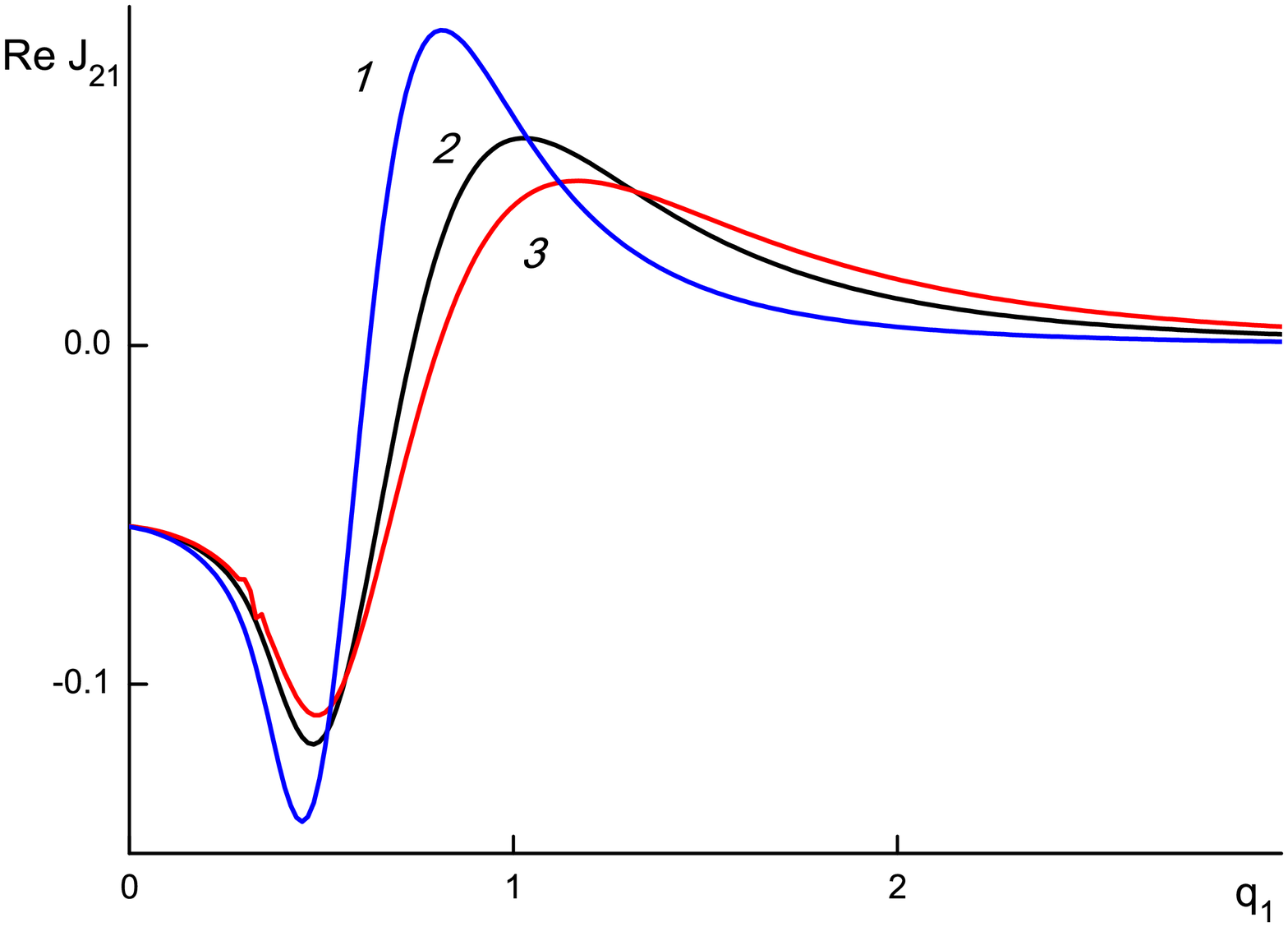}
\center{Fig. 2. Real part of dimensionless density of longitudinal
"crossed"\, current $J_{21}$,  $\Omega_1=1, \Omega_2=0.5, q_2=0.1$.
Curves $1,2,3$ correspond to values of
dimensionless chemical potential $\alpha=2, 0, -5$.}
\end{figure}

\begin{figure}[ht]\center
\includegraphics[width=16.0cm, height=10cm]{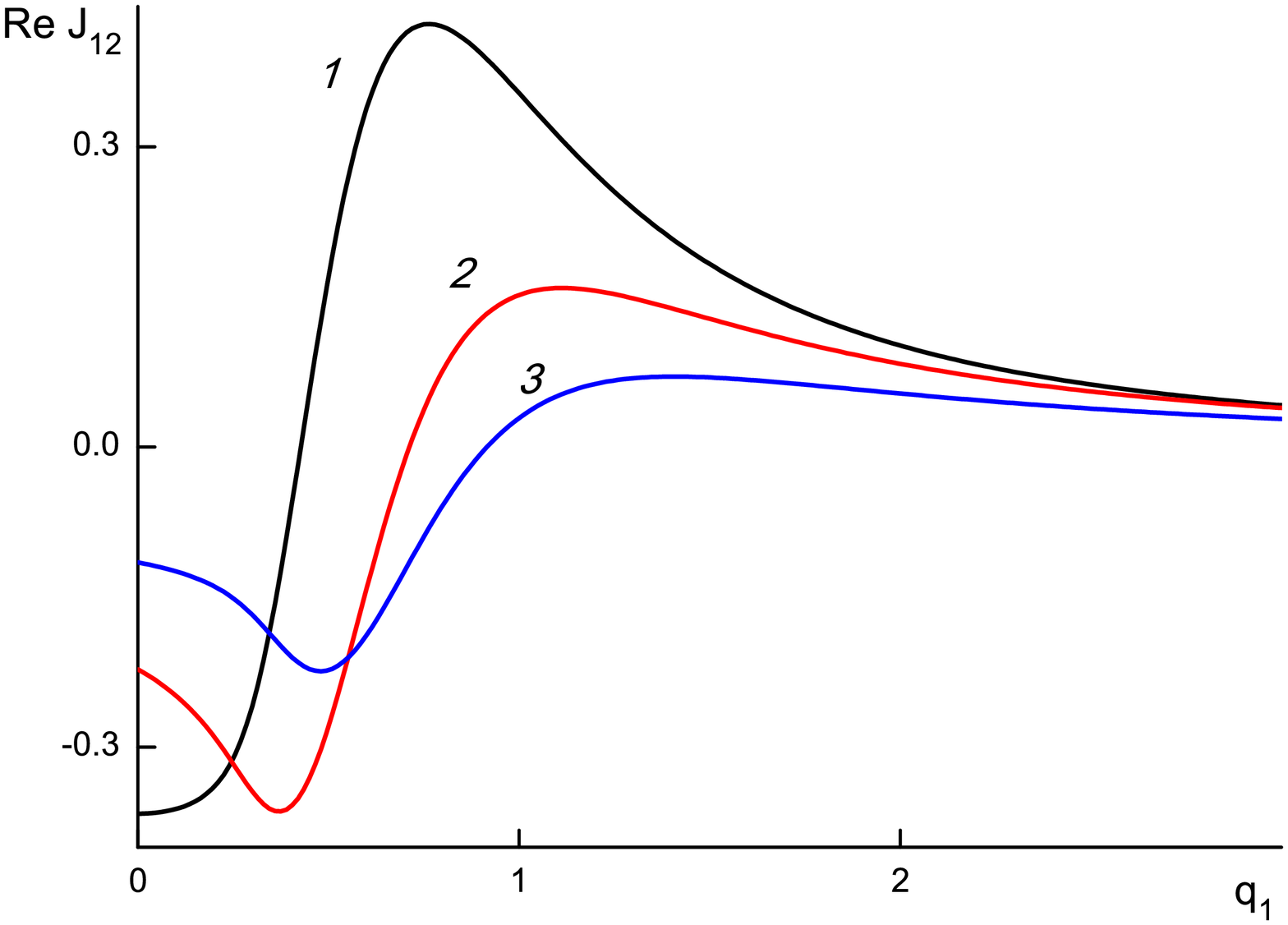}
\center{Fig. 3. Real part of dimensionless density of longitudinal
"crossed"\, current $J_{12}$, $\Omega_1=1, q_2=0.1$.
Curves $1,2,3$ correspond to values of
dimensionless oscillation frequency of second electromagnetic field
$\Omega_2=0.1, 0.2, 0.3$.}
\includegraphics[width=17.0cm, height=10cm]{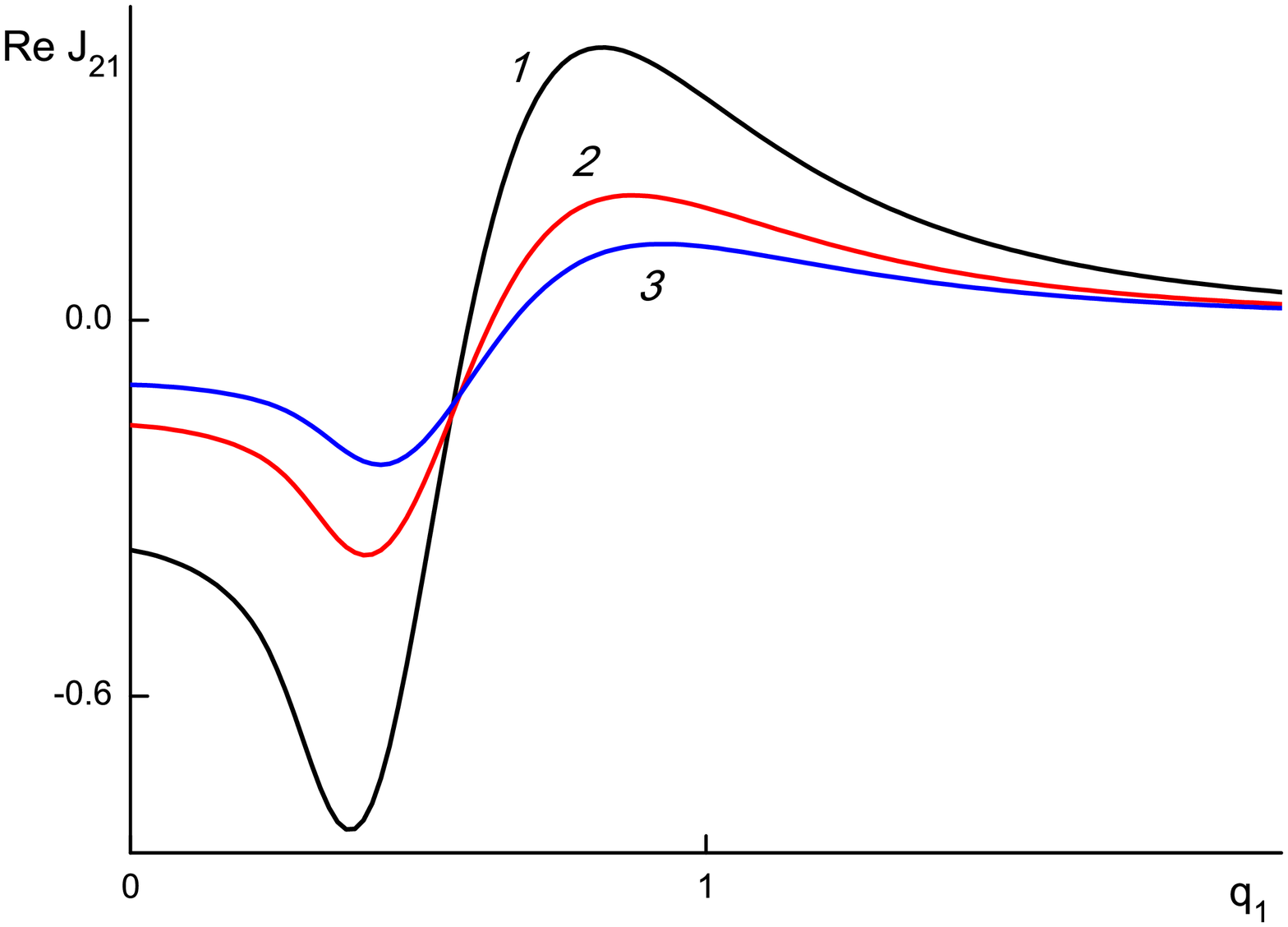}
\center{Fig. 4. Real part of dimensionless density of longitudinal
"crossed"\, current $J_{21}$, $\Omega_1=1, q_2=0.1$.
Curves $1,2,3$ correspond to values of
dimensionless oscillation frequency of second electromagnetic field
$\Omega_2=0.1, 0.2, 0.3$.}
\end{figure}

\begin{figure}[ht]\center
\includegraphics[width=16.0cm, height=10cm]{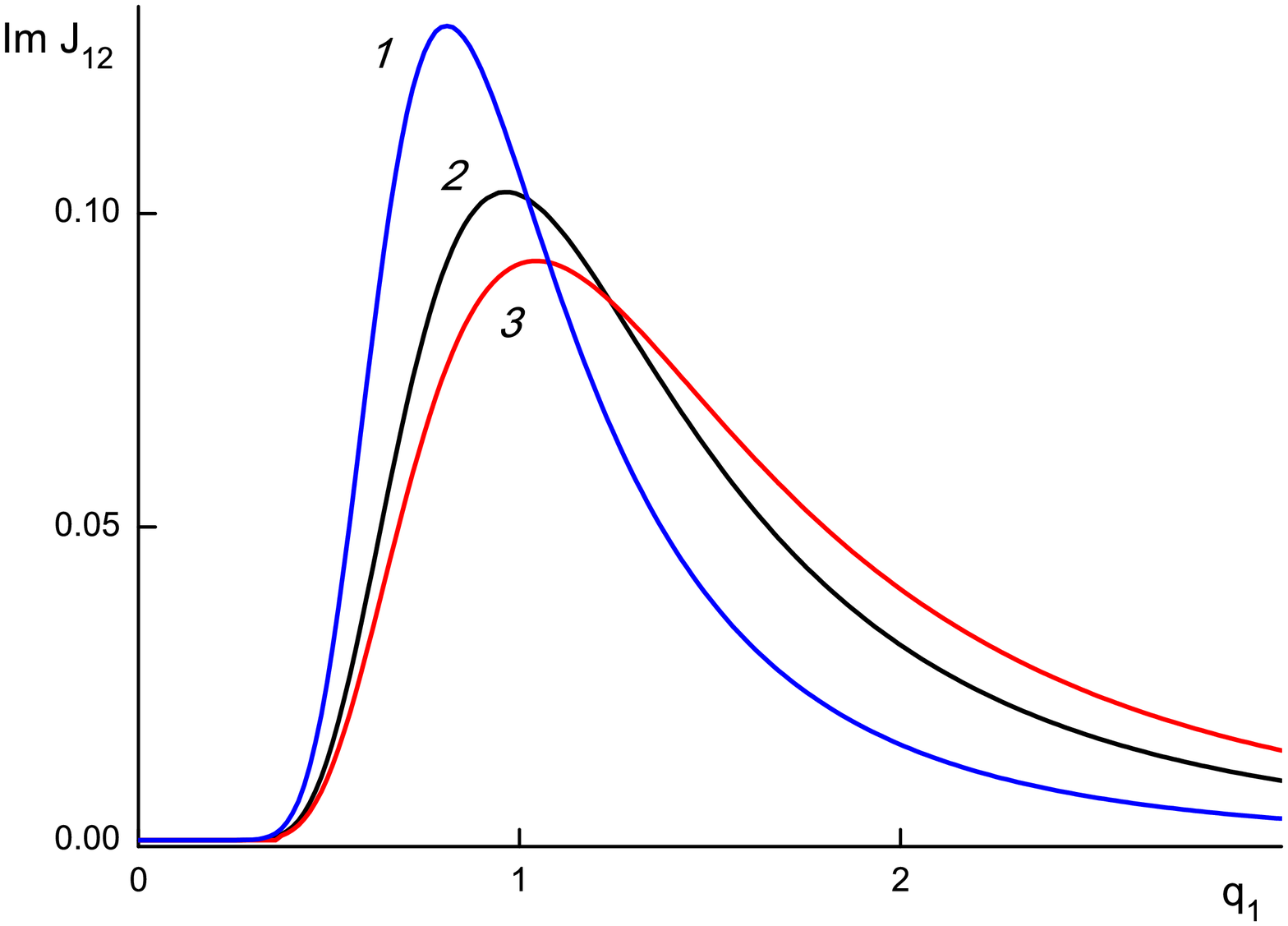}
\center{Fig. 5. Imaginary  part of dimensionless density of longitudinal
"crossed"\, current $J_{12}$, $\Omega_1=1, \Omega_2=0.5, q_2=0.1$.
Curves $1,2,3$ correspond to values of
dimensionless chemical potential $\alpha=2, 0, -5$.}
\includegraphics[width=17.0cm, height=10cm]{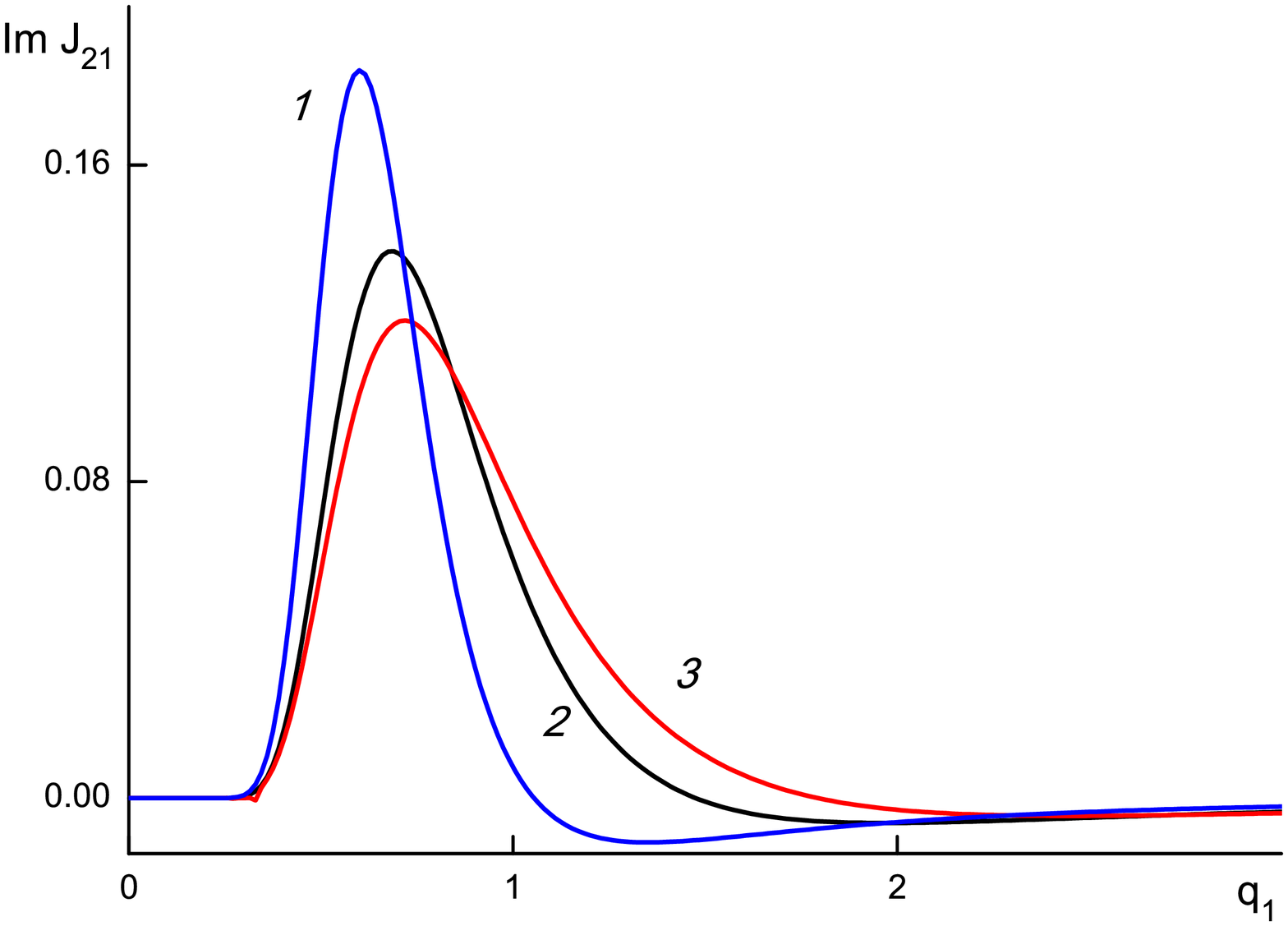}
\center{Fig. 6. Imaginary  part of dimensionless density of longitudinal
"crossed"\, current $J_{21}$, $\Omega_1=1, \Omega_2=0.5, q_2=0.1$.
Curves $1,2,3$ correspond to values of
dimensionless chemical potential $\alpha=2, 0, -5$.}
\end{figure}

\begin{figure}[ht]\center
\includegraphics[width=16.0cm, height=10cm]{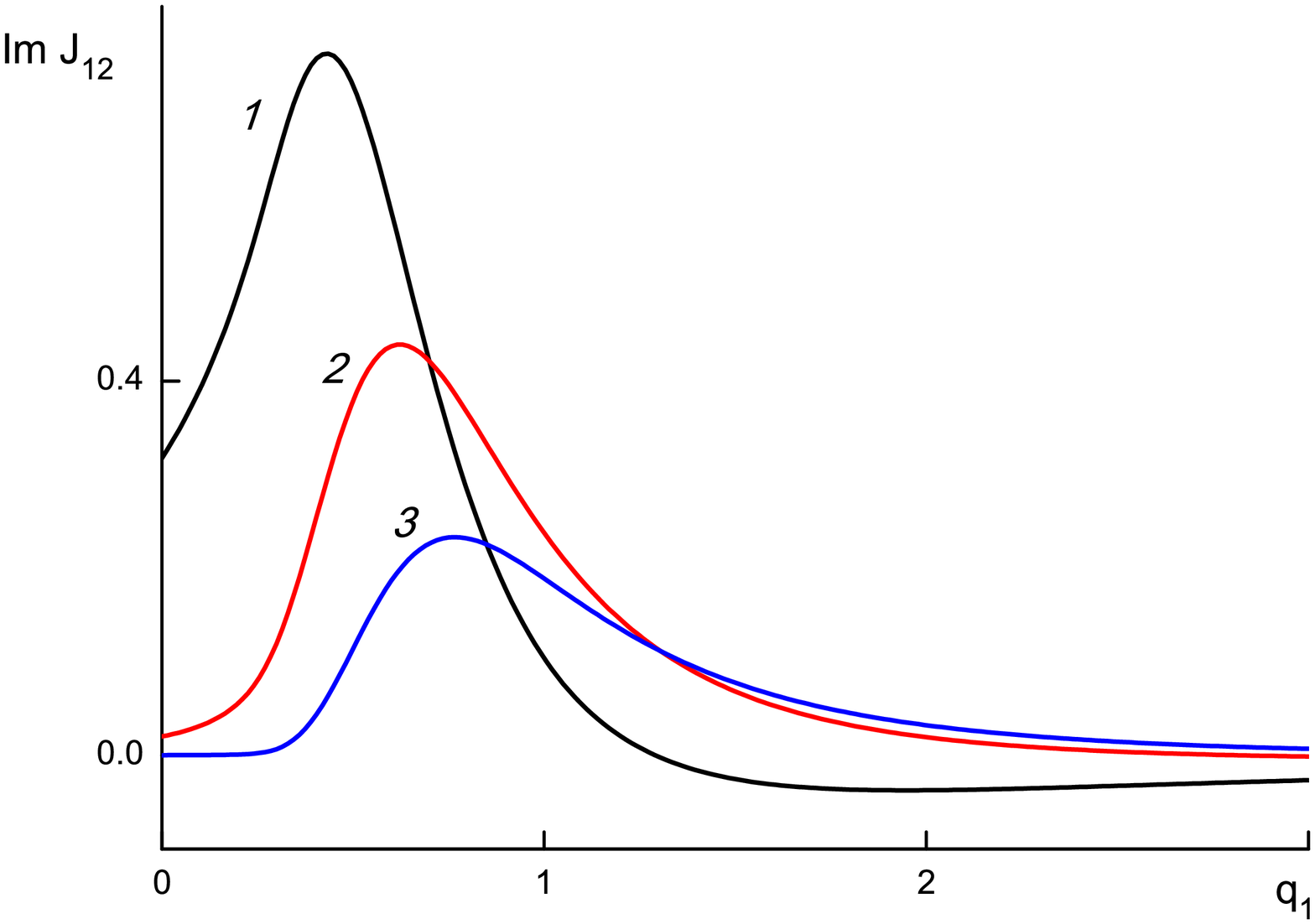}
\center{Fig. 7. Imaginary  part of dimensionless density of longitudinal
"crossed"\, current $J_{12}$, $\Omega_1=1, q_2=0.1$.
Curves $1,2,3$ correspond to values of
dimensionless oscillation frequency of second electromagnetic field
$\Omega_2=0.1, 0.2, 0.3$.}
\includegraphics[width=17.0cm, height=10cm]{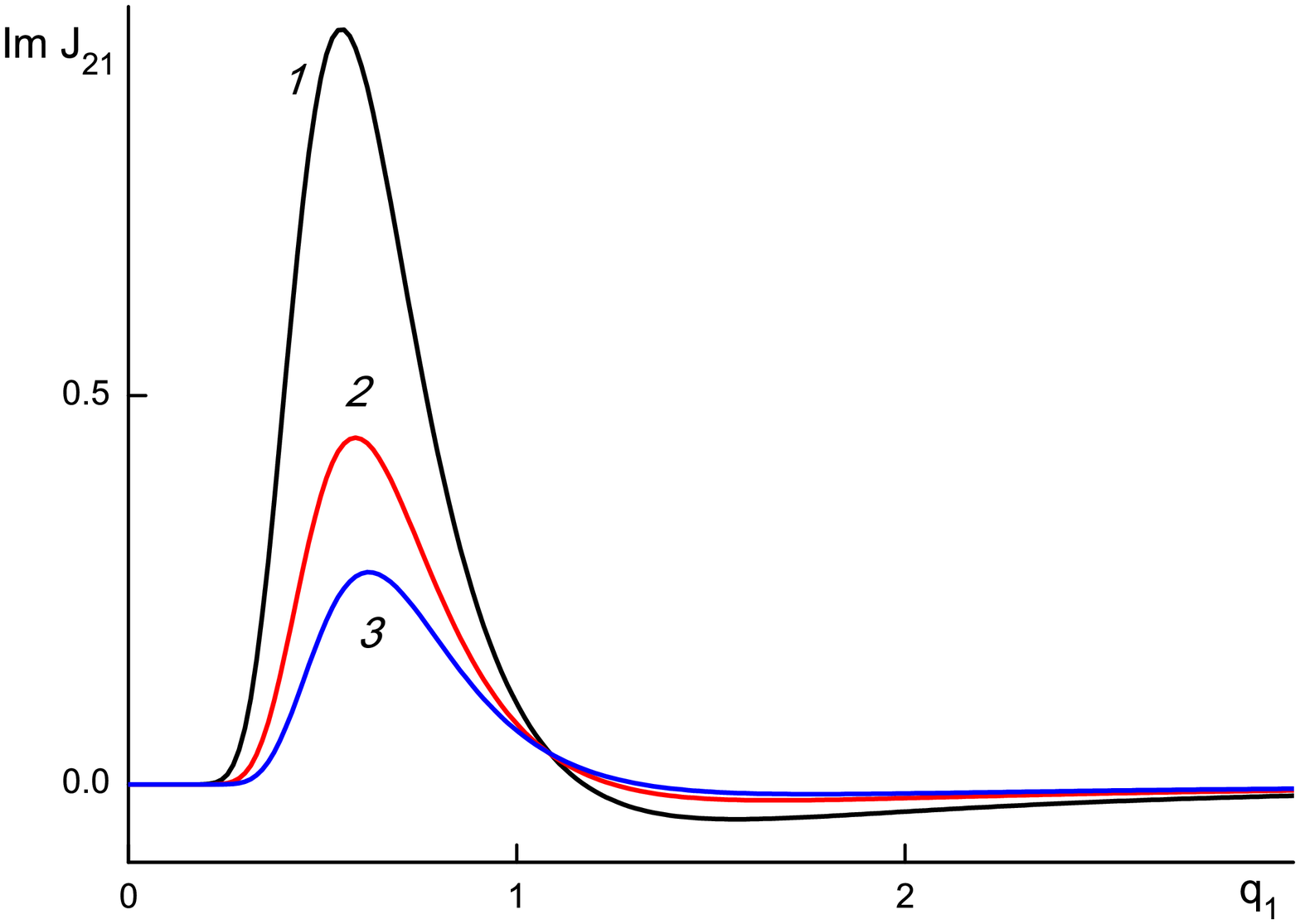}
\center{Fig. 8. Imaginary  part of dimensionless density of longitudinal
"crossed"\, current $J_{21}$, $\Omega_1=1, q_2=0.1$.
Curves $1,2,3$ correspond to values of
dimensionless oscillation frequency of second electromagnetic field
$\Omega_2=0.1, 0.2, 0.3$.}
\end{figure}

\end{document}